\documentclass[usenatbib]{mn2e}  
\usepackage{graphicx}
\usepackage{txfonts}
\usepackage{multirow}
\usepackage{supertabular}
\usepackage{longtable}
\usepackage{rotating}
\usepackage{amssymb}
\usepackage{dcolumn}

\usepackage{natbib}

\DeclareMathAlphabet{\mathsc}{OT1}{cmr}{m}{sc}
\def\testbx{bx}%
\DeclareRobustCommand{\ion}[2]{%
\relax\ifmmode
\ifx\testbx\f@series
{\mathbf{#1\,\mathsc{#2}}}\else
{\mathrm{#1\,\mathsc{#2}}}\fi
\else\textup{#1\,{\mdseries\textsc{#2}}}%
\fi}

\newcommand{\SiIV}{\ion{Si}{iv}}

\newcommand{\feii}{\ion{Fe}{ii}}

\newcommand{\mgii}{\ion{Mg}{ii}}

\newcommand{\oiii}{\ion{O}{iii}}

\def\h2{$\rm H_2$}

\def\kms{km\,s$^{-1}$}

\def\daa {$\Delta \alpha/\alpha$}

\begin{document}

\title[Constraining $\alpha$ using QSO spectra]
{Constraining the Variation in Fine-Structure Constant Using SDSS DR8 QSO Spectra}
\author[Rahmani et al.]{H. Rahmani$^{1,2}$, N. Maheshwari$^{3}$ and R. Srianand$^{1}$\\
$^{1}$ Inter-University Centre for Astronomy and Astrophysics, Post Bag 4,  Ganeshkhind, Pune 411\,007, India \\
$^{2}$ School of Astronomy, Institute for Research in Fundamental Sciences (IPM), PO Box 19395-5531, Tehran, Iran \\
$^{3}$ Indian Institute of Technology Bombay, Powal, Mumbai - 400076, India \\  
}
\pagerange{\pageref{firstpage}--\pageref{lastpage}} \pubyear{2013}
\maketitle
\label{firstpage}

\begin {abstract}
{
We report a robust constrain on the possible variation of fine-structure constant, $\alpha \equiv e^2/\hbar c$, obtained  
using \oiii\ $\lambda\lambda$ 4959,5007 nebular emission lines from QSOs.
We find  a $\Delta\alpha/\alpha=-(2.1\pm1.6)\times10^{-5}$ based on a well selected  
sample of 2347 QSOs  from Sloan Digital Sky Survey Data Release 8 with 0.02 $< z <$ 0.74. 
Our result is consistent with a non-varying $\alpha$ at a level of $2\times10^{-5}$ over approximately  7 Gyr. 
This is the largest sample of extragalactic objects yet used to constrain the variation of $\alpha$. 
While this constraint is not as stringent as those determined using many-multiplet method it  
is free from various systematic effects. 
A factor of $\sim$ 4 improvement in \daa\ achieved here compared to the previous study \citep{Bahcall04} is just consistent 
with what is expected based on a factor of 14 times bigger sample used here. This suggests that errors 
are mainly dominated by the statistical uncertainty.
We also find the ratio of transition probabilities corresponding to the \oiii\ $\lambda$5007 and $\lambda$4959 lines 
to be 2.933$\pm$0.002, in good agreement with the National Institute of Standards and Technology measurements.
}
\end{abstract}
\section{Introduction}
Most of the physical theories rely on a set of fundamental constants (e.g. 
fine-structure constant, $\alpha$ = $e^2/\hbar c$, proton-to-electron mass ratio, $\mu$, etc.) that can not be 
calculated theoretically and have to be measured experimentally. 
However, unified  theories of particle interaction like string theory  
suggest the spatial and temporal variation of these fundamental constants \citep[see][]{Uzan03,Uzan11}. 
Most of the laboratory measurements are consistent with the no  variation of 
physical constants over time-scales of $\lesssim 100$ yr \citep[e.g.][]{Rosenband08,Guena12}. 
For example, the constancy of $\alpha$ has been established  via extremely accurate laboratory 
measurements extending over 16 years resulting in  $\dot{\alpha}/\alpha <$ 10$^{-16}$ yr$^{-1}$ \citep[][]{Guena12}. 
The study of geological samples have also shown a non-varying physical 
constants over time-scales of two billion years \citep[e.g.][]{Petrov06}. 
{ Spectra of high-$z$ QSOs, in principle allow one to probe 
possible variations of dimensionless fundamental constants over 
cosmological scales.}
%

Initial attempts to measure $\alpha$ at high redshifts were based on 
the relative separation of Alkali-Doublet (AD) lines  
\citep{Savedoff56,Bahcall67b,Wolfe76,Levshakov94,Varshalovich96,Cowie95,Varshalovich00,Murphy01, Chand05}. 
\citet{Chand05} used a sample of 23 \SiIV\ absorbers, observed with 
Very Large Telescope Ultraviolet and Visual Echelle Spectrograph (VLT/UVES),
to find \daa\footnote{Here \daa\ is defined as $\left(\alpha_z-\alpha_0\right)/\alpha_0$ 
where $\alpha_z$  and $\alpha_0$ are the measured values of $\alpha$ at any redshift, $z$, and in 
the laboratory on the Earth.} = $-(0.02\pm0.55)\times 10^{-5}$ 
which is the best constraint 
on \daa\ based on AD method.
%
Higher sensitivities in \daa\ ($\lesssim 10^{-5}$) can be achieved using   
Many-Multiplet (MM) method in which one 
simultaneously correlates different multiplets from several ions \citep{Dzuba99a,Dzuba99b,Webb99}.
\citet{Murphy03} applied the MM technique on a sample of 128 QSO absorbers observed with 
High Resolution Echelle Spectrometer (HIRES) on Keck to find a \daa\ = $-5.7\pm1.0$ ppm 
 which shows $\alpha$ is smaller at higher redshifts. On the  contrary, the analysis of  
a VLT/UVES sample of 21 \mgii\ systems by  \citet{Srianand07} resulted in  a  \daa\ = $+0.1\pm1.5$ ppm,         
consistent with a no  variation in $\alpha$ at high redshifts. 
{ Null results are also obtained using only Fe~{\sc ii} multiplets
of few individual systems \citep[][]{Quast04,Chand06,Levshakov07}.
}
\citet{Webb11} compiled a large sample 
of QSOs from both Keck/HIRES and VLT/UVES to claim a spatially varying $\alpha$ with a dipole pattern. 
This claim is not yet verified independently \citep[see for example,][]{Molaro13}. 
Although using MM method one reaches high sensitivities in \daa\ 
{ it is possible that this method may suffer from
systematics related to ionization and chemical homogeneities.
In addition it has also been found that different high resolution
spectroscopic data used  presently suffer from
large and small scale wavelength calibration errors \citep{Griest10,Whitmore10,Rahmani13}.
Therefore, it is important to have independent measurements using
different instruments and measurement techniques.}
Stringent constraint on fundamental constants can be obtained by comparing the 21-cm redshift with that of UV lines. 
Applying such a techniques on  a sample of 
four \mgii\ absorbers \citet{Rahmani12} found a \daa\ = $0.0\pm1.5$ ppm, consistent with no variation in $\alpha$. 
The major uncertainty in this technique comes from the difficulties in associating 
the 21-cm component with the corresponding UV absorption line component. 
\oiii\  $\lambda\lambda4959,5007$ are two strong nebular emissions, with  a doublet separation of 
$\Delta\lambda_{\rm OIII} = 47.9320$ \AA,  
seen in the spectrum of most of the QSOs and star-forming galaxies. 
A comparison between the laboratory value of $\Delta\lambda_{\rm OIII}$ and its value measured 
from a QSO  leads to 
a constraint on \daa\  in the range of $10^{-4}$--$10^{-3}$.
\citet{Bahcall04} applied such a technique on 165 well selected QSO spectra 
published by Sloan Digital Sky Survey (SDSS) Data Release one (DR1) 
to find \daa\ = $+(1.2\pm0.7)\times10^{-4}$. In this work, we apply the same technique to a much larger sample 
of QSOs available  in SDSS DR8 \citep[][]{Aihara11} to obtain a stringent constrain on the value of $\alpha$. 
%
In contrary to absorption line techniques, the effect of systematic errors will 
be minimized due to  the large sample of available QSOs. 
Star-forming galaxies are also suitable for such studies as they have narrow \oiii\ emission lines that 
are hardly contaminated by broad H$\beta$ emission as frequently 
seen in QSOs. However, we have chosen QSOs as they spread over much larger redshifts than galaxies 
and also have a well defined power-law continuum that makes the analysis 
using automated procedures more straightforward. 
Furthermore, intrinsic emission line profiles of galaxies are not usually resolved in the SDSS  spectra. This makes 
the estimate of the line centroids to be dominated by the  systematic errors. 
This paper is organized as follows. 
In section \ref{sample_qso} we explain our 
sample of QSO. We present our algorithm for measuring \daa\ from each QSO in section \ref{cr-analysis}.
Results and conclusions are presented in section  \ref{results} and \ref{conclusion}, respectively. 
\section{QSO Sample}\label{sample_qso}
The QSO sample used in this study comes from the spectroscopic sample of QSOs 
published by SDSS DR8 
\citep{Aihara11}. 
We begin with a sub-sample of SDSS DR8 QSOs with  $ z \le$ 0.74. At  $z \gtrsim 0.74$, the 
\oiii\ doublet falls at the observed wavelength of 
$\gtrsim$ 8712 \AA\ where the  SDSS spectrum is usually filled with lots 
of spikes most likely due to residuals from  subtraction of strong sky emission lines. As our 
exercise requires very high quality  data we have excluded QSOs with their \oiii\ 
emission in these regions.  
{ There are  26368 QSOs within the redshift range considered above.}
We further notice that 
a significant fraction of QSOs have poor spectral quality close to \oiii\ emission lines that 
can lead to highly unreliable \daa\ measurements. It is important to exclude such 
systems from our analysis. By trying different filters we found that the following set of conditions 
can confidently reject the majority  of such QSOs:
(i) The amplitude of \oiii\ $\lambda4959$ emission, $A_1$, must be larger than five times of the average error; 
(ii) The amplitude-ratio  of \oiii\ $\lambda5007$ to \oiii\ $\lambda4959$ emission, $A_2/A_1$, must be 
greater than 1. Ideally, $A_2/A_1 \sim 3$; (iii) There should not be any pixel with bad flag { in wavelength range of \oiii\ lines};  
(iv) The \oiii\ doublets should not be so broad that their profiles overlap. We implement this by considering 
those doublet where $5\sigma < (\lambda_2-\lambda_1)/2$ where $\sigma$ is the width of the best fitted Gaussian 
to \oiii\ emissions.
The preliminary cuts are very modest to remove only the worst spectra. The remaining 12016 QSO spectra can 
still have various problems which makes them not ideally suited for \daa\ measurements. 
We now apply additional selection filters suggested by \citet{Bahcall04} to further prune our sample. 
\subsection{Signal-to-noise ratio of \oiii\ emission}
To have precise measurements we need a very clean detection of \oiii\ emission lines. \oiii\ 
doublets with poor SNR can lead to \daa\ measurements with large systematic errors. 
To choose QSO spectra with clean \oiii\ emission lines we accept only those
QSOs having \oiii\ fluxes detected with a 
SNR of at least 15. Here we calculate the noise from the scatter of the flux in the line free region 
used to fit the continuum in the vicinity of the \oiii\ emission lines. This cut leaves us with 8721 QSOs. 
\subsection{Broad H$\beta$ emission} 
H$\beta$ $\lambda$4861 line is the closest emission line to the \oiii\ $\lambda$4959 line. It is very well 
known that H$\beta$ emission is usually broad. 
A very broad  
H$\beta$ line, which is frequently seen in QSOs spectra, can distort the emission profile 
of \oiii\ $\lambda$4959 and can lead to wrong $\alpha$ measurements. 
We require to find a condition based on which we can check if the emission profile of 
H$\beta$  has significant overlap with the  \oiii\ $\lambda$4959 profile. 
To do so we only accept QSOs that pass the following two conditions:
(i) equivalent width (EW) of H$\beta$ is two time smaller than 
the EW of \oiii\ $\lambda$5007; 
(ii) fraction of H$\beta$ flux that overlaps with \oiii\ $\lambda$4959 
to be less than 2\%. 
Only 4707 out of 8721 QSOs pass through such a filter.
\begin{figure} 
\hspace*{-.6cm}	
\includegraphics[width=0.99\hsize,bb=18 18 594 774,clip=,angle=90]{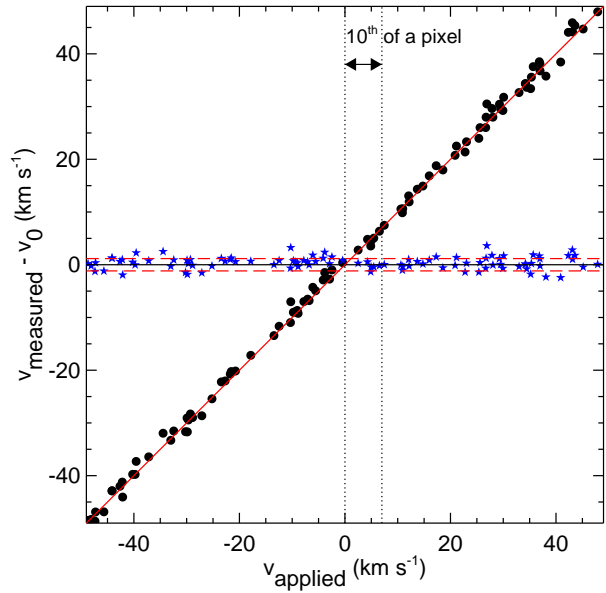}
\caption{
Result of  simulations to check the accuracy of  our cross-correlation analysis.
The abscissa is the applied shift and the ordinate is the mean of the measured 
shifts for 100 realizations. On the solid line the measured and applied shifts are equal. 
The asterisks are the residuals (measured - applied) and the long dashed lines are the 
mean and 1$\sigma$ scatter of the residuals. The two vertical dashed lines 
indicate  1/10$^{th}$ of a pixel size ($\Delta$v $\sim$ 7 \kms).
}
\label{fig_sim}
\end{figure}
\subsection{Kolmogorov-Smirnov Test}
The estimated value of \daa\ is very sensitive to the shape of the \oiii\ doublet emission profiles. Therefore, any mismatch 
between the shapes of the doublet emissions { (due to unknown contamination)} can lead to a 
wrong \daa\ measurement. Here we make use of a seven point Kolmogorov-Smirov (KS) test 
to quantify the similarity between the shapes of the two \oiii\ emission lines. 
To do so we determine whether the flux values in seven pixels centered on the \oiii\ $\lambda$4959 emission are  drawn from 
the same distribution as those of \oiii\ $\lambda$5007.  We require that the two sets to 
be drawn from the same distribution with 95\% confidence level (corresponding to 2$\sigma$). 
Only 2428 of the remaining 4707 QSOs pass this test. 
\subsection{Narrow \oiii\ emission line}  
The resolution power of SDSS spectra is $\sim$ 2000 which is sampled approximately 
by three pixels of sizes $\sim$ 70 \kms. The \oiii\ emission should be 
well resolved out of the SDSS resolution to have well defined intrinsic line shape. 
Therefore, we reject QSOs with very narrow \oiii\ emissions where 
their 2$\sigma$ width of the \oiii\ lines are less than 200 \kms. This condition is very mild (in comparison to other cuts) to reduce 
the number of QSOs from 2428 to 2347.
\begin{figure} 
\hspace*{-0.6cm}	
\vbox{
\includegraphics[width=0.8\hsize,bb=18 18 594 774,clip=,angle=90]{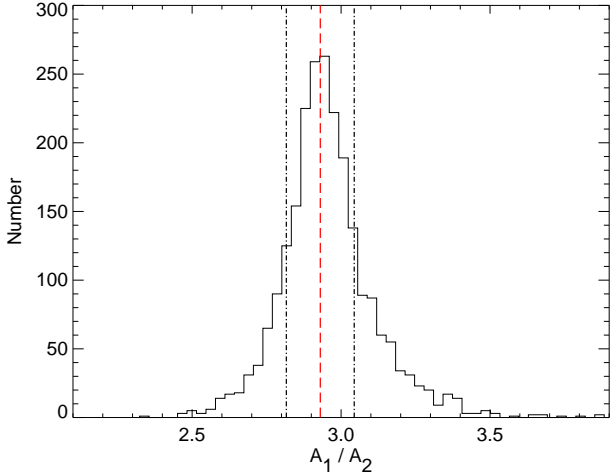}
}
\caption{Histogram of the measured amplitude ratios of the [\oiii] doublet for our final sample 
of QSOs. The weighted mean, shown as long-dashed line, corresponds to 2.933$\pm$0.002. The 
vertical dashed-dotted lines presents  the weighted standard deviation of the measured values.
%
%
}
\label{fig_hist}
\end{figure}
\begin{figure} 
\hspace*{-.8cm}	
\includegraphics[width=0.9\hsize,bb=18 18 594 774,clip=,angle=90]{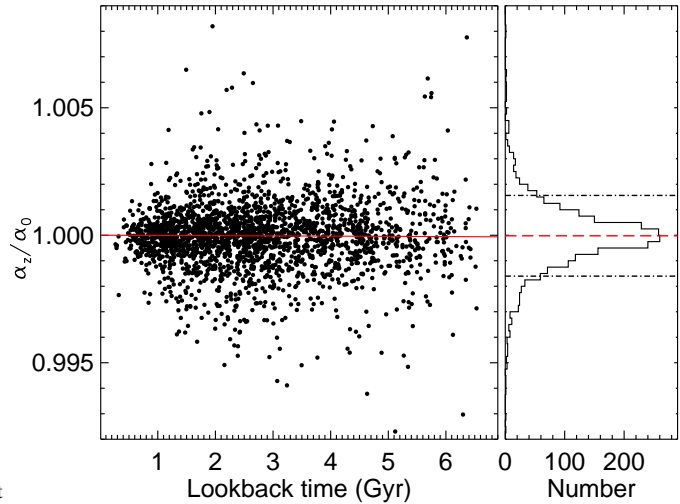}
\caption{{\it Left: } the values of fine-structure constant compared to its laboratory 
value, $\alpha_z/\alpha_0$, vs lookback time. The red solid 
line presents the best fitted line with a slope of $-(0.9\pm1.1)\times10^{-5}$ and 
the intercept of $\alpha_z/\alpha_0 - 1$ = $(0.1\pm1.1)\times10^{-5}$.  
{\it Right}: the histogram of $\alpha_z/\alpha_0$. The long-dashed line presents the weighted mean 
($\alpha_z/\alpha_0 - 1$ = $-2.1\times 10^{-5}$ )
and the dashed-dotted lines present the 2$\sigma$ range ($\sigma = 0.00079$) where $\sigma$ 
is the weighted standard deviation. 
}
\label{fig_alpha_t}
\end{figure}
The collection of above cuts defines our "final" sample of 2347 QSOs. We will present \daa\ measurements 
for this sample based on a cross correlation analysis.  
\subsection{\feii\ emission lines}
\feii\ $\lambda$4923 and \feii\ $\lambda$5018 are two \feii\ emission lines that are sometimes seen in the 
spectra of QSOs in the vicinity of \oiii\ lines. Such a close emission line can influence our 
measurements as they can distort the shape of the \oiii\ emission lines. 
However, as predicted by \citet{Bahcall04} KS test ensures such contamination are not sever in our sample. 
Inspecting dozens of randomly chosen spectra from our final sample, we did not find any of the 
QSOs having the above \feii\ emissions. { We further stacked spectra of all QSOs in our final sample  
and did not detect any of these \feii\ emissions in the stacked spectrum.} 
Therefore, such \feii\ emissions will have negligible effect in our \daa\ measurements and can not 
bias our results. 

Even though we have used Gaussian fits to define our sample from the full SDSS data set, we use 
cross-correlation techniques (described below) to measure \daa. 
\section{cross-correlation analysis for \daa\ measurements}\label{cr-analysis}
The main step in measuring $\alpha$ from a QSO spectrum is to estimate $\Delta\lambda_{\oiii}$, the 
wavelength difference between the two \oiii\ doublet emissions. By further comparison of $\Delta\lambda_{\oiii}$ and 
its laboratory value, 47.9320 \AA, we will express one \daa\ for each QSO. 
Cross-correlation analysis has been frequently used for estimating the velocity offset 
between similar spectral features in the literature \citep[See][for examples]{Wendt11,Agafonova11,Rahmani12,Rahmani13}. 
Here, we elaborate a cross-correlation analysis to estimate  $\Delta\lambda$. 
To do so we shift each spectrum to the rest frame of the QSO and convert the scales from wavelength to velocity. 
We then rebin the spectra into new pixel arrays of sizes 10 \kms\ using a cubic spline interpolation. Finally we 
perform a cross-correlation analysis between the two \oiii\ emissions which is expressed as following
\begin{equation}
h(V) = (f\star g)(V)=\int_{-\infty}^{\infty} f(v)g(V+v)\,\mathrm{d} v
\end{equation}
where $f(v)$ and $g(v)$ correspond to the \oiii\ emission lines which are functions of 
velocity, $v$, and $h(V)$ is the cross-correlation function 
where $V$ is the shift. The function $h(V)$  peaks at a velocity, $V_0$, where the two 
\oiii\ doublet profiles best match. We estimate the $V_0$ as the peak of a Gaussian 
function fitted to $h(V)$. The value of fine structure constant at the redshift 
of the QSO, $\alpha(z)$, can then be estimated as 
\begin{equation}
\Delta \alpha / \alpha \equiv \frac{\alpha(z)}{\alpha(0)} - 1 =  \sqrt{1+\frac{V_0}{2c\Lambda_0^2}} - 1, ~~~~~\Lambda_0 = \sqrt{\frac{\lambda_2-\lambda_1}{\lambda_2+\lambda_1}}
\label{eq_alpha}
\end{equation}
where $c$ is the speed of light and $\lambda_1$ and $\lambda_2$ are the laboratory wavelengths 
of the \oiii\ doublet emission lines.  Hence by measuring the $V_0$ we directly estimate a \daa\ based on 
each QSO spectrum. We further follow a Monte Carlo simulation to associate a statistical error to 
each measured \daa. To do this we first generate 100 random realizations of our original QSO spectrum  
using its error spectrum. We then calculate a \daa\ for each of the realized spectra following exactly the 
same procedure as that of the original spectrum. Finally we calculate the standard deviation of these 100 estimated \daa\ 
and quote it as 1$\sigma$ error of \daa. 

The most important step in estimating a \daa\ from a QSO spectrum is measuring $V_0$.  
Any systematic error in our cross-correlation analysis can leave biases in our results and 
lead to unreliable conclusions. Hence it becomes utmost important to check our 
cross-correlation against any kind of systematic error. To do so we perform a simulation 
analysis as following: (1) we first measure the velocity shift  $\rm v_0$ for  a randomly 
chosen QSO. (2) We then apply a velocity shift, $\rm v_{applied}$, to this spectrum and 
generate 100 realization spectra from this shifted spectrum using its error spectrum. (3) Making use of 
our cross-correlation routine we measure the velocity shift for each of the 100 realizations to obtain 
the mean shift of $\rm v_{measured}$. (4) Finally we repeat such an exercise for a sample of applied shifts 
in the range of   $-$50 -- 50 \kms. Fig. \ref{fig_sim} presents the results of this analysis. 
Clearly the residual differences  between the applied and the  measured shifts are randomly distributed 
around zero with a scatter of  smaller than tenth of a pixel size. As a result we exclude the possibility 
that our final values of $\alpha$ is  affected by some systematics related to our procedure of measuring shifts.
%
%
\begin{table}
\caption{\daa\ for various sub-samples of our final sample of QSOs.}
\begin{center}
\begin{tabular}{lccc}
\hline
Sub-sample$^\dagger$  & Sample size & \multicolumn{2}{c}{\daa\ (10$^{-5}$)} \\
                     &             & weighted mean & simple mean$^\star$ \\
\hline
$z < 0.21$                &   1164      & $-2.4\pm2.1$  & $-0.6\pm1.7$ \\
$z > 0.21$                &   1164      & $-1.7\pm2.6$  & $-1.4\pm2.6$ \\
$\sigma < 3.4$ \AA        &   1164      & $-2.2\pm2.2$  & $-1.7\pm1.7$ \\
$\sigma > 3.4$ \AA        &   1164      & $-2.0\pm2.7$  & $+1.7\pm2.6$ \\
SNR      $ < 38.5$        &   1164      & $-12.4\pm3.8$ & $-6.5\pm3.1$ \\
SNR      $ > 38.5$        &   1164      & $+0.3\pm1.8$  & $+0.5\pm1.5$ \\
\hline
\end{tabular}
\end{center}
\begin{flushleft}
$^\dagger$ All sub-samples are made based on the median of the given parameters in this column  that 
are standing for $z$ of the QSO, best fitted $\sigma$ to \oiii\ profile, 
and the SNR of \oiii\ $\lambda$4959.\\
$^\star$ Simple mean after 2$\sigma$ clipping. 
\end{flushleft}
\label{tab_sub}
\end{table}
\section{results}\label{results}
In this section we summarize the results we get based on the analysis of our final 
QSO sample. 
Fig. \ref{fig_hist} presents the distribution of the amplitude 
ratios of the two \oiii\ doublet lines, $A_2/A_1$. The distribution 
has a mean of 2.933$\pm$0.002 which is in agreement with its best theoretically estimated value, 2.92, from 
National Institute of Standards and Technology (NIST) Atomic Spectra Database \citep[][]{Wiese1996}. We would like to recall that 
this ratio is calculated based on our best fitted Gaussian profiles to \oiii\ doublets. Such an  
agreement shows that our profile fitting procedure works very well. This is an important issue as 
the majority of the filters we have defined are built based on the Gaussian profile fitting.  
Fig. \ref{fig_alpha_t} in its {\it left panel} presents our measured $\alpha(z) / \alpha(0)$ vs the 
lookback time. We have estimated the lookback time based on a standard $\Lambda$CDM 
background cosmology \citep{Hinshaw09} for the redshift of the QSOs. 
Our best fitted line to these points shows a slope of 
$(-0.9\pm1.1)\times10^{-5}$ and an 
intercept of \daa\ = $(0.1\pm1.1)\times10^{-5}$ that are consistent with a no variation 
in fine-structure constant over last 7 Gyr. 
The histogram of estimated $\alpha(z) / \alpha(0)$ is shown in the 
{\it right panel} of Fig. \ref{fig_alpha_t}. We find a weighted mean of  $-(2.1\pm1.6)\times10^{-5}$ 
with a weighted standard deviation of 0.00079 for our measured \daa. The reduced $\chi^2$ for the 
weighted mean is 1.1 which shows the quoted error is acceptable. However, we also estimate a 
simple mean after rejecting outliers by a 2$\sigma$ clipping to get \daa = $-(1.9\pm1.5)\times10^{-5}$ 
with a standard deviation of $\sigma = 0.00061$. 
The estimated weighted mean and simple mean are consistent with each other and with a no variation 
in the fine-structure constant within 2$\sigma$ errors. Furthermore, the evaluated weighted and standard 
errors are very much consistent which shows our estimated errors for individual \daa\ are realistic. 
Clearly these measurements provide a substantial improvement to \daa\ = $+(1.2\pm0.7)\times10^{-4}$ 
found by \citet{Bahcall04}. 
One of the main issues in \daa\ measurements is the wavelength stability. 
Fitting sky and arc lines for each fiber to find the wavelength solution has led to a quite good 
spectroscopic wavelength calibration in SDSS DR7 and later releases.  The typical wavelength calibration error  
reaches 2 \kms\ and can be still less in the red part of the spectrograph \citep{Abazajian09}. 
By inserting a $V_0 = 2$ \kms\ in Eq. \ref{eq_alpha} we convert such an error to (\daa)$_{\rm cal}$ = $3\times10^{-4}$. 
The typical statistical error of \daa\ measurements in our study is $\sim$ $10\times10^{-4}$, which 
is 3 times larger than (\daa)$_{\rm cal}$. In addition, we   
expect such calibration errors act randomly over a large sample of objects.  
We further notice that the two spectrograph of SDSS disperse the incoming light on 
two CCDs called {\it blue} and {\rm red} where the former covers from 3900--6100 \AA\ and the latter 
from 5900--9100 \AA. Hence, a wavelength range of 5900--6100 \AA\ of each object is covered by two 
spectrograph. Such an overlap with two possible different wavelength solutions in the edges of the two CCDs can impact our results. 
To check such an effect we exclude those QSOs having their \oiii\ emissions in the above mentioned range 
from our final sample of QSOs. However, for the remaining (1983) QSOs 
we find \daa\ =    $-(1.7\pm1.7)\times10^{-5}$ for the weighted mean and  \daa\ =    $-(2.1\pm1.6)\times10^{-5}$ 
for the simple mean after 2$\sigma$ clipping which are consistent with the results we obtained from 
our final sample of QSOs. Therefore, our results are not affected by the "possible" systematics due to 
the different wavelength solutions in the overlapping regions of the two CCDs.  

In Table \ref{tab_sub} we have further explored the value of \daa\ for some more 
sub-samples of our final sample of QSOs. We have divided our final sample of QSOs into two parts 
based on the median values of respectively  $z$ of the QSOs, $\sigma$ of the best fitted Gaussian to \oiii\ lines, and 
the SNR of the total flux of the \oiii\ $\lambda$4959 lines.  
We present both the weighted mean and the simple mean after 2$\sigma$ clipping 
for all sub-samples. Interestingly there exists a reasonable  match between the 
two estimated errors for each sub-sample. 
This is a signature for the correct estimate of the error of individual \daa\ measurements. 
The low SNR sub-sample is the only case that 
is consistent with more than 2$\sigma$  variation of $\alpha$ while 
having the largest measured error as well. Other sub-samples are always consistent 
with a stable $\alpha$ with no variation. 
As expected better constraints are obtained in high SNR and narrow  albeit resolved 
emission lines sub-samples. 
\section{Conclusion}\label{conclusion}
We have made use of an appropriately chosen sub-sample of QSOs in SDSS DR8 
to constrain the possible variation of fine-structure constant 
by using the \oiii\ $\lambda\lambda$ 4959,5007 nebular emission lines. 
Our final sample of QSOs consists of 2347 objects. This is the largest sample of objects 
yet used for constraining the variation of constants. 
We find \daa\ =  $-(2.1\pm1.6)\times10^{-5}$ at the mean redshift of   $z \sim 0.2$.  This  
is consistent with  a no  variation of $\alpha$ over last 7 Gyr with an accuracy of 10 part in million.  
This is roughly a factor four improvement compared to the existing measurements based on \oiii\ doublets \citep{Bahcall04}. 
However, this constraint is an order of magnitude weaker than those obtained 
from MM method \citep{Murphy03,Srianand07}. However, because of the 
large sample of objects and the simplicity of the method our result is   
much less affected by the systematic errors due to inhomogeneities in the absorbing medium and wavelength 
calibration errors. Furthermore, we find that our estimated \daa\ is fairly consistent in different sub-samples 
of our main sample of QSOs. As a byproduct of our analysis, we estimated the amplitude ratio of \oiii\ doublet 
to be 2.933$\pm$0.002 which is in an excellent agreement with its theoretically predicted value, 2.92,  from NIST.

\citet{Bahcall04} had analysed the same \oiii\ doublets from 165 QSOs 
chosen from SDSS DR1  to find \daa\ =  $+(1.2\pm0.7)\times10^{-4}$. 
Having a sample that is $\sim$ 14 times larger than that of \citet{Bahcall04}, one expects 
to reach an accuracy of $\sim 0.7\times10^{-4}/14^{0.5} = 1.9 \times 10^{-5}$. This is very 
close to what we have achieved in our current study. 
This also illustrate that a 100 fold increase in QSO spectra (i.e. $\sim 10^{5}$) is required 
to reach the sensitivity of one parts per million in \daa\ using \oiii\ doublets. 

\section*{acknowledgments}
We acknowledge the use of SDSS spectra from the archive (http://www.sdss.org/). 
NM wishes to thank the Indian Academy of Science for their support through Summer Research 
Fellowship Programme 2011. 
\def\aj{AJ}%
\def\actaa{Acta Astron.}%
\def\araa{ARA\&A}%
\def\apj{ApJ}%
\def\apjl{ApJ}%
\def\apjs{ApJS}%
\def\ao{Appl.~Opt.}%
\def\apss{Ap\&SS}%
\def\aap{A\&A}%
\def\aapr{A\&A~Rev.}%
\def\aaps{A\&AS}%
\def\azh{AZh}%
\def\baas{BAAS}%
\def\bac{Bull. astr. Inst. Czechosl.}%
\def\caa{Chinese Astron. Astrophys.}%
\def\cjaa{Chinese J. Astron. Astrophys.}%
\def\icarus{Icarus}%
\def\jcap{J. Cosmology Astropart. Phys.}%
\def\jrasc{JRASC}%
\def\mnras{MNRAS}%
\def\memras{MmRAS}%
\def\na{New A}%
\def\nar{New A Rev.}%
\def\pasa{PASA}%
\def\pra{Phys.~Rev.~A}%
\def\prb{Phys.~Rev.~B}%
\def\prc{Phys.~Rev.~C}%
\def\prd{Phys.~Rev.~D}%
\def\pre{Phys.~Rev.~E}%
\def\prl{Phys.~Rev.~Lett.}%
\def\pasp{PASP}%
\def\pasj{PASJ}%
\def\qjras{QJRAS}%
\def\rmxaa{Rev. Mexicana Astron. Astrofis.}%
\def\skytel{S\&T}%
\def\solphys{Sol.~Phys.}%
\def\sovast{Soviet~Ast.}%
\def\ssr{Space~Sci.~Rev.}%
\def\zap{ZAp}%
\def\nat{Nature}%
\def\iaucirc{IAU~Circ.}%
\def\aplett{Astrophys.~Lett.}%
\def\apspr{Astrophys.~Space~Phys.~Res.}%
\def\bain{Bull.~Astron.~Inst.~Netherlands}%
\def\fcp{Fund.~Cosmic~Phys.}%
\def\gca{Geochim.~Cosmochim.~Acta}%
\def\grl{Geophys.~Res.~Lett.}%
\def\jcp{J.~Chem.~Phys.}%
\def\jgr{J.~Geophys.~Res.}%
\def\jqsrt{J.~Quant.~Spec.~Radiat.~Transf.}%
\def\memsai{Mem.~Soc.~Astron.~Italiana}%
\def\nphysa{Nucl.~Phys.~A}%
\def\physrep{Phys.~Rep.}%
\def\physscr{Phys.~Scr}%
\def\planss{Planet.~Space~Sci.}%
\def\procspie{Proc.~SPIE}%
\let\astap=\aap
\let\apjlett=\apjl
\let\apjsupp=\apjs
\let\applopt=\ao
\bibliographystyle{mn}
\bibliography{mybib}
\appendix

\end{document}